\documentclass[11pt,onecolumn,english,showpacs,preprintnumbers,amsmath,amssymb,floatfix,nofootinbib]{revtex4-1}

\usepackage[T1]{fontenc}
\usepackage[latin9]{inputenc}
\usepackage{color}
\usepackage{array}
\usepackage{amstext}
\usepackage{graphicx}
\usepackage{esint}
\usepackage{cleveref}
\usepackage{rotating}
\usepackage{slashed}
\usepackage[normalem]{ulem}
\usepackage{siunitx}
\usepackage[font=small,labelfont=bf]{caption}
\usepackage{xcolor}
\usepackage[colorlinks = true,
linkcolor = magenta,
urlcolor  = blue,
citecolor = red,
anchorcolor = blue]{hyperref}

\makeatletter

\@ifundefined{textcolor}{}
{%
	\definecolor{BLACK}{gray}{0}
	\definecolor{WHITE}{gray}{1}
	\definecolor{RED}{rgb}{1,0,0}
	\definecolor{GREEN}{rgb}{0,1,0}
	\definecolor{BLUE}{rgb}{0,0,1}
	\definecolor{CYAN}{cmyk}{1,0,0,0}
	\definecolor{MAGENTA}{cmyk}{0,1,0,0}
	\definecolor{YELLOW}{cmyk}{0,0,1,0}
}

\@ifundefined{definecolor}
{\usepackage{color}}{}
\@ifundefined{definecolor}
{\usepackage{color}}{}
\makeatother
\usepackage{babel}

\begin{document}
	
	
\title { Constraining top quark flavor violation and dipole moments through three and four top quark productions at the LHC 	  }

\author{Malihe Malekhosseini$^{1}$}
\email{M\_Malekhosseini@semnan.ac.ir}

\author{Mehrdad Ghominejad$^{1}$}
\email{Mghominejad@semnan.ac.ir}

\author{Hamzeh Khanpour$^{2,3}$}
\email{Hamzeh.Khanpour@mail.ipm.ir}

\author{Mojtaba Mohammadi Najafabadi$^{3}$}
\email{Mojtaba@cern.ch}

\affiliation {
$^{(1)}$Faculty of Physics, Semnan University, Semnan P.O.Box 35131-19111, Semnan, Iran  \\
$^{(2)}$Department of Physics, University of Science and Technology of Mazandaran, P.O.Box 48518-78195, Behshahr, Iran \\
$^{(3)}$School of Particles and Accelerators, Institute for Research in Fundamental Sciences (IPM), P.O.Box 19395-5531, Tehran, Iran  }

\date{\today}

%
\begin{abstract}\label{abstract}

In this paper, we examine the sensitivity of the three-top quark production at the LHC to the top quark 
flavor-changing neutral currents (FCNC) as well as the sensitivity of the four-top production to the strong and weak dipole 
moments of the top quark.  Upper limits at $95\%$ CL on the branching fractions of $\mathcal{B}(t\rightarrow qX)$, where $X = g, Z, \gamma, H$ and $q=u,c$, are set
by performing an analysis on three-top events in the same-sign dilepton channel. We consider the main sources of the background processes and 
a realistic detector simulation is performed at the center-of-mass energy of 14 TeV. 
In the second part of this work, based on the recent upper limits which have been set on the four-top quark 
cross section by the ATLAS and CMS collaborations from 13 TeV data, 
we constrain the top quark strong and weak dipole moments.  The bounds on the 
top quark dipole moments are presented using the future LHC prospects for four-top quark cross section measurement.

\end{abstract}
%


\maketitle


%
\section{Introduction}\label{sec:intro}

In proton-proton collisions at the LHC, mostly top quarks are produced in pair via strong interaction~\cite{Sirunyan:2016cdg,CMS:2017pkf} 
or singly via weak interactions~\cite{Sirunyan:2016cdg,CMS:2017pkf}.
However, the large center-of-mass energy of proton-proton collisions at the LHC opens the possibility to have three~\cite{Barger:2010uw,Chen:2014ewl}  
or four-top quark productions.
Searches for four-top quark production using the Run I dataset at $\sqrt{s} = 8 \, {\rm TeV}$ have been performed by both the CMS~\cite{Khachatryan:2014sca,Chatrchyan:2013fea} and ATLAS~\cite{Aad:2015kqa,Aad:2015gdg} experiments, with no observed excess of data above the background expectation.
The searches of both experiments have been updated using the 13 TeV data using different final states~\cite{Sirunyan:2017roi,Sirunyan:2017tep,Beck:2016hyi,ATLAS:2016gqb,Alvarez:2016nrz,CMS:2017wvz,CMS:2016wig}. 
The cross section of the four-top quark production at the LHC with the center-of-mass energy of 13 TeV is around $\sigma(pp \rightarrow t \bar{t} t\bar{t}) = 9$ fb \cite{x1,x2}.
Within the SM at leading-order, the three-top quarks are produced in association with either a $W$ boson or a jet with a total rate of around 
2 fb ~\cite{Barger:2010uw,Chen:2014ewl}.
Although three and four-top rates are extremely small with respect to the $t \bar{t}$~\cite{Aaboud:2017fha,Aaboud:2016syx} or single top
 quark production by around five order of magnitudes,
these processes are particularly sensitive to New Physics (NP) beyond the Standard Model (SM). 
Beyond the SM scenarios predict enhancements in $t t \bar{t} X$ and $t \bar{t}t \bar{t}$ production cross section
~\cite{Barger:2010uw,Chen:2014ewl,Cacciapaglia:2011kz,AguilarSaavedra:2011ck,Kanemura:2015nza,Deandrea:2014raa,Bevilacqua:2012em,Han:2012qu,Gregoire:2011ka}. Vector-like quarks, SUSY with R-parity violation are examples of the BSM scenarios which affect their rates and some have been experimentally studied~\cite{Khachatryan:2014sca,Aad:2015kqa,ATLAS:2016gqb,CMS:2016wig,ATLAS:2016sno,Aad:2016tuk,CMS:2016ahn}.

So far, the top quark is the heaviest discovered particle with its mass close to the scale of electroweak symmetry breaking and Yukawa coupling near one, $y_{t} \sim 1$. 
Therefore, one may expect that possible NP effects would show up in top quark production or decays
~\cite{Husemann:2017eka,Aguilar-Saavedra:2014iga,Barducci:2017ddn,Schulze:2016qas,Buckley:2015lku,Arkani-Hamed:2015vfh,Guo:2016kea,Lillie:2007hd,Cortiana:2015rca,Boos:2015bta,Cristinziani:2016vif,Escamilla:2017pvd}. NP can be seen either directly in new particles production or through the indirect effects via the higher order corrections. Indeed, the observation of indirect
indications is important because it provides hints to search for new physics before direct observation.  Within the 
SM framework, the branching fractions of the rare decays of the top quark $t \rightarrow qX$, with 
$X = g, \gamma, Z$, Higgs and $q=u,c$, are extremely small and are of the order of $10^{-14}-10^{-12}$~\cite{Agashe:2013hma}.
Due to smallness of these branching fractions, the current and future experiments would not be able to measure
them. Such transitions in the SM are only possible at loop-level and are significantly suppressed because of the Glashow-Iliopoulos-Maiani (GIM) mechanism~\cite{Glashow:1970gm}.
On the other part, it has been found that many SM extensions potentially can relax the suppression in 
top quark FCNC decays from GIM mechanism leading to considerable enhancements for
$\mathcal{B}(t\rightarrow qX)$~\cite{Agashe:2013hma}. This happens because of the appearance of several loop diagrams with new particles mediated inside them.
Beyond the SM scenarios like technicolor, SUSY models, two-Higgs doublet models predict much higher branching fractions of the order of $\sim10^{-10}$ to  $10^{-6}$ which are 
larger than the SM values~\cite{Agashe:2013hma,Bejar:2008ub,Cao:2008vk,Arhrib:2005nx,Gaitan:2015hga,Yang:2013lpa,DiazCruz:2009ek,Cao:2007dk,Dey:2016cve}.
Searches for FCNC in the top quark sector have been followed by various experiments in the past years, including ALEPH, DELPHI,
L3 and OPAL experiments at LEP~\cite{Abbiendi:2001wk,Achard:2002vv,Abdallah:2003wf,Heister:2002xv}, 
H1 and ZEUS experiments at HERA~\cite{Moretti:1997dz,Chekanov:2003yt,Aktas:2003yd,Abramowicz:2011tv,Aaron:2009vv}, 
CDF and D0 experiments at the Tevatron~\cite{Abe:1997fz,Aaltonen:2008ac,Abazov:2011qf,Abazov:2007ev,Aaltonen:2008qr}, and CMS and ATLAS experiments at the LHC. 
The most stringent limits on the top FCNC branching fractions come from the LHC experiments.
The CMS collaboration presented the result of their search for the FCNC through single-top-quark production in association with a photon~\cite{Khachatryan:2015att}.
Based on an integrated luminosity of 19.8 fb$^{-1}$ in $pp$ collisions at a center-of-mass energy of 8 TeV,
upper limits on $\mathcal{B} (t \rightarrow q \gamma)$ and $\mathcal{B} (t \rightarrow q Z)$ 
through FCNC single-top-quark production in association with a photon or a $Z$ boson are obtained~\cite{Khachatryan:2015att, Sirunyan:2017kkr}.
The upper limits on the branching fractions at $95\%$ CL are: 
\begin{eqnarray}
&&\mathcal{B}(t \to u \gamma) < 1.3 \times 10^{-4} ~,~\mathcal{B}(t \to c \gamma) < 1.70 \times 10^{-3} \,,     \\
&&\mathcal{B}(t \to u Z) < 2.2 \times 10^{-4} ~,~\mathcal{B}(t \to c Z) < 4.9 \times 10^{-4} \,.
\end{eqnarray}
The most recent search for the $tqH$ FCNC has been done by looking at the events  with either single top quark FCNC
production associated with a Higgs boson and $t\bar{t}$ production with FCNC decay of 
one of the top quarks~\cite{Aaboud:2017mfd,CMS:2017cck,Aad:2015pja}. The search is based on the Higgs boson decay into a $b\bar{b}$ pair and uses the data  corresponding
to an integrated luminosity of 35.9 fb$^{-1}$ at $\sqrt{s} = $ 13 TeV.  The
observed upper limits at $95\%$ CL on the branching fractions of FCNC top quark decays are~\cite{CMS:2017cck}:
\begin{eqnarray}
\mathcal{B}(t\rightarrow uH)< 4.7 \times 10^{-3} ~,~ \mathcal{B}(t\rightarrow cH)< 4.7\times 10^{-3}\,.
\end{eqnarray}
These upper limits are the most stringent ones to date. Among all the FCNC top quark decays, the upper limits on $\mathcal{B}(t \rightarrow q g)$  
has been tightly bounded. Using the data collected at the center-of-mass energy of 8 TeV corresponding 
to an integrated luminosity of 20.3 fb$^{-1}$ by the ATLAS detector, the observed upper limits on the branching fractions are found to be~\cite{Aad:2015gea}:
\begin{eqnarray}
\mathcal{B}(t \rightarrow u g) < 4.0 \times 10 ^{-5} ~,~ \mathcal{B}(t \rightarrow c g) < 2.0 \times 10 ^{-5} \,.
\end{eqnarray}
The CMS experiment search for the FCNC $tqg$ provides the following limits at the $95\%$ CL~\cite{Khachatryan:2016sib}:
\begin{eqnarray}
\mathcal{B}(t \rightarrow u g) < 2.0 \times 10 ^{-5} ~,~\mathcal{B}(t \rightarrow c g) < 4.1 \times 10 ^{-4} \,. 
\end{eqnarray}
As it can be seen, the ATLAS experiment has obtained stronger bound on $\mathcal{B}(t \rightarrow c g)$
than the CMS experiment and vice versa for the $\mathcal{B}(t \rightarrow u g)$. So far, there have been a lot of studies on the top quark FCNC in proton-proton, 
electron-positron and electron-proton colliders~\cite{Khatibi:2014via,Khatibi:2015aal,Etesami:2010ig,Durieux:2014xla,TurkCakir:2017rvu,Denizli:2017cfx,Senol:2012fc,Cakir:2010rs,Khanpour:2014xla,Hesari:2015oya,Aguilar-Saavedra:2017vka,Shen:2018mlj}.
In the first part of this work, we perform a search for the FCNC interactions separately at the $tuX$
and $tcX$ vertices at the LHC by looking for events with only three-top quark in the final state.
We focus on the decay channels in which two same-sign top quarks decay into a W boson and a
$b$-quark, followed by the W boson decay to an electron or muon and a neutrino. 
The same-sign dilepton decay channels are considered because it has a  clean signature and does not suffer from large background contribution.

In the second part of the work, we study the strong and electroweak dipole moments of the top quark. 
In the SM, at leading order dipole interactions do not exist, however, 
electroweak radiative corrections generate both strong and weak dipole moments for the top quark. 
The size of top quark dipole moments are so small that the LHC would not be able to observe them. 
But, there are several well-motivated beyond the SM theories which contribute to these
dipoles and lead to sizable enhancements which make the dipoles accessible by the LHC detectors~\cite{Ibrahim:2010hv,Ibrahim:2011im,Yang:2013ula}.
As a consequence, any observation of considerable deviations from zero would be an indication to beyond the SM physics. 
Here, we investigate the sensitivity of the four-top quark production to the weak and strong top quark dipole moments 
and upper limits are set on them  using the present and prospects upper limits on the four-top quark production cross section.

The present article is organized as follows: In Section~\ref{sec:framework}, we present the theoretical framework for 
describing the top quark FCNC couplings and dipole moments which is based on the effective field theory approach.
The sensitivity estimations of three-top quark production on FCNC couplings are given in Section~\ref{Sensitivity-FCNC}. 
Section~\ref{Sensitivity-dipole} is dedicated to probe the strong and weak dipole moments using the four-top quark production.
Finally, Section~\ref{sec:Discussion} is dedicated to summarize the results and conclusions.

%
\section{Theoretical framework and assumptions}\label{sec:framework}
%

In this paper, we assume that new physics effects in three-top quark and four-top-quark productions
 are not going to be directly discovered at the LHC (like new heavy degrees of freedom).
These kinds of effects could be described by an effective Lagrangian with the following form~\cite{AguilarSaavedra:2008zc,AguilarSaavedra:2009mx}:

\begin{eqnarray}
\mathcal{L}_{\rm eff} = \mathcal{L}_{\rm SM}+\frac{1}{\Lambda^{2}}\sum_{i}c_{i}O^{d=6}_{i} + h.c.,
\end{eqnarray}

where $c_{i}$ coefficients are dimensionless by which the new physics effects are parameterized with dimension-six
operators $O_{i}$. The scale of new physics is denoted by $\Lambda$ and the $O_{i}$ are a complete set of dimension-six operators
which satisfy the SM symmetries, i.e. the Lorentz and the $SU(3)\times SU(2)\times U(1)$ gauge symmetries.
Due to the observed 
excellent agreement between the predications of the SM and data, the 
deviations from the SM are expected to be small. 

There are different dimension-six operators which contribute to the three-top and four-top-quark productions at the LHC. 
For instance, the impacts of full set of four-fermion $qqtt$ dimension-six operators on the  four-top
production have been studied in Ref.~\cite{Zhang:2017mls}.
In this work, the aim is to probe the FCNC interactions ($tqg, tqZ, tq\gamma, tqH$) coming from dimension-six operators via
three-top-quark production and the effects of dimension-six operators on $gt\bar{t}$ and $Zt\bar{t}$ couplings through the four-top-quark production at the LHC.

Following on the notation presented in Ref.~\cite{AguilarSaavedra:2008zc}, the operators $,O^{23(32)}_{uG\phi},O^{23(32)}_{uW},O^{23(32)}_{uB\phi},
O^{23(32)}_{u\phi}$ contribute to the FCNC couplings and $O^{33}_{uG\phi}, O^{33}_{uB\phi},O^{33}_{uW}$ and $O^{33}_{\phi u}$
give contributions to $gt\bar{t}$ and $Zt\bar{t}$ interactions.
The most general effective Lagrangian describing the FCNC interactions can be parameterized as:

\begin{eqnarray}
\label{Eff-Lagrangian}
\mathcal{L}_{\text{FCNC}}&=& \sum_{q = u, c} \bigg[\frac{g_s}
{2m_{t}}\bar{q}\lambda^a \sigma^{\mu \nu}
(\zeta_{qt}^L P^L+\zeta_{qt}^R P^R) t G^a_{\mu \nu}
-\frac{1}{\sqrt{2}}\bar{q}(\eta_{qt}^L P^L+\eta_{qt}^R P^R) t H  \nonumber  \\
&-&\frac{g_W}{2 c_W}\bar{q}\gamma^\mu (X_{qt}^L P_L+X_{qt}^R P_R) t Z_\mu   
+\frac{g_W}{4 c_W m_Z} \bar{q}\sigma^{\mu \nu} (\kappa_{qt}^L P_L
+\kappa_{qt}^R P_R) t Z_{\mu \nu}     \nonumber     \\
&+&\frac{e}{2 m_t} \bar{q} \sigma^{\mu \nu}(\lambda_{qt}^L P_L
+\lambda_{qt}^R P_R)t A_{\mu \nu} \bigg] + \text{h.c.},
\end{eqnarray}

where $\zeta_{qt}$, $\eta_{qt}$, $X_{qt}$, $\kappa_{qt}$ and $\lambda_{qt}$ are the real 
parameters which determine the strength of FCNC interactions with gluon, Higgs, $Z$ and photon, respectively. 
At tree-level, in the SM, all the above coefficients are zero, i.e. $\zeta_{qt}^L = \zeta_{qt}^R = \eta_{qt}^L = \eta_{qt}^R=
X_{qt}^L = X_{qt}^R  = \kappa_{qt}^L= \kappa_{qt}^R  = \lambda_{qt}^L = \lambda_{qt}^R = 0.0$.
After the electroweak symmetry breaking, these FCNC parameters are related to the Wilson coefficients of dimension six operators \cite{AguilarSaavedra:2008zc}.
For example, the FCNC $tqg$ parameters $\zeta_{qt}^L$  and $\zeta_{qt}^R$ are related to the Wilson coefficients 
through:
\begin{eqnarray}
\zeta_{qt}^L =  \frac{\sqrt{2}\text{Re}(C^{32*}_{uG\phi}) v m_{t}}{g_{s}\Lambda^{2}} ~~,~~ \zeta_{qt}^R= \frac{\sqrt{2}\text{Im}(C^{23}_{uG\phi}) v m_{t}}{g_{s}\Lambda^{2}}
\end{eqnarray}

The effective Lagrangian including dimension-six operators for $gt\bar{t}$ and $Zt\bar{t}$
can be written as:

\begin{eqnarray}\label{Lag:gttbar}
\mathcal{L}_{g t \bar{t}} =
- g_s \bar{t} \frac{\lambda^a}{2} \gamma^{\mu} t G^{a}_{\mu}
- g_s \bar{t} \lambda^{a} \frac{i \sigma^{\mu \nu} q_{\nu}}{m_t}
(d_{V}^{g} + i d_{A}^{g} \gamma_5) t G^{a}_{\mu}, 
\end{eqnarray}
and
\begin{eqnarray}
\mathcal{L}_{Zt\bar{t}}  = -\frac{g_{W}}{2 c_{W}} \bar{t}
\gamma^{\mu}\left( X_{tt}^{L} P_{L}
+ X_{tt}^{R} P_{R} - 2 s_{W}^{2} Q_{t}\right) t \ Z_{\mu}  
-\frac{g_{W}}{2 c_{W}} \bar{t} \frac{i \sigma^{\mu \nu}
q_{\nu}}{m_{Z}}\left(d_{V}^{Z} + i \ d_{A}^{Z}\gamma_5 \right) t \ Z_{\mu},
\end{eqnarray}

The couplings $d_{V}^{g(Z)}$ and $d_{A}^{g(Z)}$ are real parameters and are related to strong (weak) magnetic and strong (weak) 
electric dipole moments of the top quark, respectively. In the SM, at tree-level, $d_{V}^{g(Z)} = 0.0$ and $d_{A}^{g(Z)} = 0.0$
while the QCD and electroweak corrections at loop level generate very tiny values for the weak and strong dipole moments~\cite{Martinez:2001qs,Martinez:2007qf, Bernabeu:1995gs, Chien:2015xha,Hollik:1998vz, Agashe:2006wa}.
The values of $X_{tt}^{L}$ and $X_{tt}^{R}$ parameters are equal to one and zero in the SM, respectively.

The most stringent bounds on $d_{V}^{g}$ and $d_{A}^{g}$ are derived from the low energy measurements.
In particular, $d_{V}^{g}$ is constrained using the rare $B$-mesons decays~\cite{Martinez:2001qs} and 
$d_{A}^{g}$ is probed through the neutron electric dipole moment ($d_{n}$)~\cite{Kamenik:2011dk}. The $95\%$ CL limits from the low energy experiments are~\cite{Martinez:2001qs,Kamenik:2011dk}:
\begin{eqnarray}
-3.8\times 10^{-3} \leq d_{V}^{g} \leq 1.2 \times 10^{-3}~,~|d_{A}^{g}| \leq 0.95\times 10^{-3},
\end{eqnarray}
The combination of the LHC and Tevatron top quark pair cross section measurements leads to the following $95\%$ CL regions~\cite{Aguilar-Saavedra:2014iga}: 
\begin{eqnarray}
-0.012 \leq  d_{V}^{g} \leq 0.023~,~  |d_{A}^{g}| \leq 0.087.
\end{eqnarray}
As it can be seen, the indirect constraints are tighter than the direct ones by one order of magnitude.

The  weak dipole moments $d^{Z}_{V}$ and $d^{Z}_{A}$  have been studied at the LHC and a future electron-positron collider using the $t\bar{t}Z$ production \cite{Rontsch:2015una}.
At the LHC, both $d^{Z}_{V}$ and $d^{Z}_{A}$ are expected to be probed down to the order of 0.15 with 300 fb$^{-1}$ of integrated luminosity of data which would be well-improved
to $0.08$ using 3000 fb$^{-1}$ integrated luminosity of data. Limits from the electroweak precision data are found to 
be at the same order. A future $e^{-}e^{+}$ collider at the center-of-mass energy of 500 GeV with 500 fb$^{-1}$ 
would be able to achieve the limits of $0.08$ on $|d^{Z}_{A}|$ and $[-0.02,0.04]$ on the $d^{Z}_{V}$~\cite{Rontsch:2015una}.

\section{Sensitivity of three-top quark production to FCNC couplings}\label{Sensitivity-FCNC}
%

\begin{figure}[htb]
\begin{center}
\vspace{0.50cm}
\resizebox{0.55\textwidth}{!}{\includegraphics{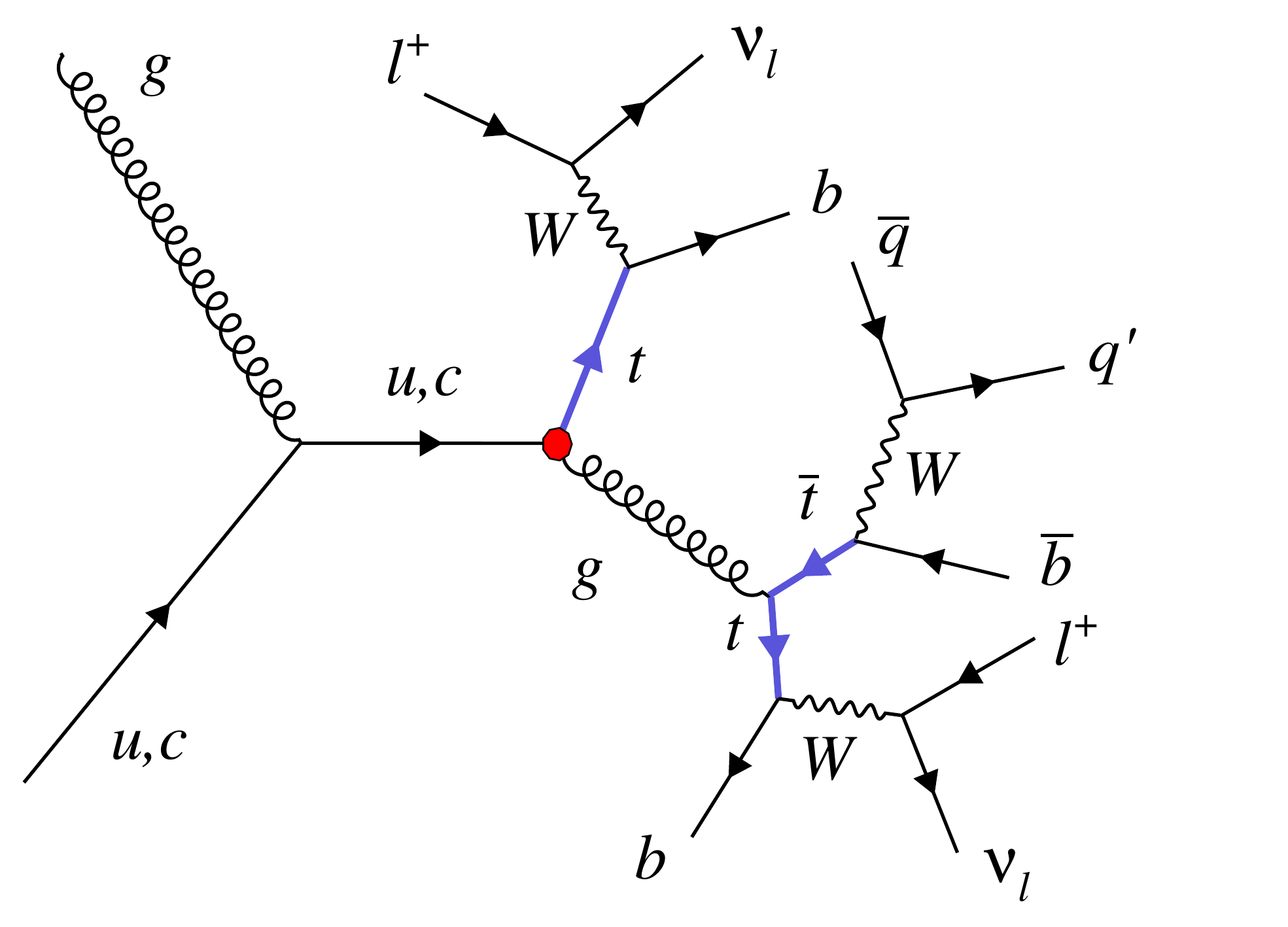}}   		
\caption{  \small {A representative leading order Feynman diagram of the $tqg$ FCNC contributions to the three-top quark production at the LHC including the lepton decay 
of the W boson from the top quark decay and hadronic decays of the W boson from the anti-top quark. 
\label{fig:Feynman-diagrams} } }
\end{center}
\end{figure}

In this section, we study the sensitivity of the three-top quark production 
cross section to the FCNC interactions of $tqg$, $tq\gamma$, $tqZ$, and $tqH$.
In the SM at leading order, similar to the single top quark production, three-top quark events 
are produced in association with a light-jet, or a b-jet, or associated with a $W$ boson, i.e.
\begin{eqnarray}
pp\rightarrow t t \bar{t} (t\bar{t}\bar{t}) + \text{jet}~,~pp\rightarrow t t \bar{t} (t\bar{t}\bar{t}) + \text{$b$-quark}~,~
pp\rightarrow t t \bar{t} (t\bar{t}\bar{t}) + W,
\end{eqnarray}
The sum of cross section of all processes amounts to around 1.9 fb at the LHC with the
center-of-mass energy of 14 TeV. The presence of FCNC couplings $tqX$, $X=g,\gamma,Z,H$,
would lead to production only three-top quark (which does not exist at leading order in SM). 
Figure \ref{fig:Feynman-diagrams} shows the lowest-order diagram for the three-top quark production from $tqg$ FCNC coupling
including the leptonic decays of the $W$ boson from top quarks and hadronic decay of the $W$ boson from 
anti-top quark.
The FCNC vertex in this Feynman diagram is denoted by a filled circle. The diagrams for the other FCNC interactions
are similar except that the gluon should be replaced by a photon, a $Z$-boson, or a Higgs boson.

In this work, the search is performed for all FCNC couplings of $tqg$, $tq\gamma$, $tqZ$, and 
$tqH$ independently, i.e. one is switched on at a time. We also do the analysis separately for $q=u$
and $q=c$. We concentrate on a very clean signature with two same-sign leptons, where lepton could be 
either an electron or a muon. Therefore, the signal events are generally characterized by the presence of 
exactly two isolated same-sign charged leptons, large missing transverse energy, and several
jets from which three of them come from $b$-quarks.  We perform the analysis for 300 and 3000 fb$^{-1}$ of the LHC  at the center-of-mass energy
of 14 TeV and present the upper limits on the branching fractions of $\mathcal{B}(t\rightarrow qX)$ at  $95\%$ CL.

%
\subsection{Event simulation and selection}
%

In this section, the simulation tools and techniques as well as the event selection and reconstruction are described. The process of signal is
taken as three-top quarks followed by the leptonic decay of two same-sign top quarks and hadronic decay of the
other top quark. As a result, the final state consists of two same-sign charged leptons, at least five jets from which three are originating from
$b$-quarks, and missing transverse energy. In this exploratory study,  the background processes such as $t\bar{t}Z$, $t\bar{t}W$, SM four-top,
$t \bar{t}WW$, $t \bar{t}ZZ$ and $WWZ$ are considered. For both signal and the background processes,  
we use \texttt{MadGraph5\_aMC@NLO} package~\cite{Alwall:2014hca}. It  automatically generates the code for obtaining the production rates.
The leading order parton distribution functions of NNPDF 3.0~\cite{Ball:2014uwa} are employed as the input for the calculations and the mass of the top quark is set to 173 GeV.
To calculate the three-top quark cross sections in the presence of the  FCNC couplings, the effective Lagrangians introduced above are implemented
into the \texttt{FeynRules} package~\cite{Alloul:2013bka,Duhr:2011se} and is exported  into a UFO module~\cite{Degrande:2011ua} which is connected to \texttt{MadGraph5\_aMC@NLO} \footnote{The UFO file is taken from http://feynrules.irmp.ucl.ac.be/wiki/GeneralFCNTop}.

To obtain the cross section of three-top versus the FCNC couplings, the calculations in the presence of 
$\zeta_{qt}$, $\lambda_{qt}$, $\kappa_{qt}$ and $\eta_{qt}$ are done  assuming various values: $\pm 1.0,\pm 2.0 $ and fit the resulting 
cross sections to quadratic polynomials.  Considering each FCNC coupling at a time, 
the cross sections times the related top quark effective coupling strengths are given in  the first column of Table~\ref{tab:cs}.

\begin{table}[tbh]
\begin{center}
\begin{tabular}{c|c|c|c|c|c}
$\sqrt {s}$ = 14 TeV     ~&~  Cross section (in fb)           ~&~  Cut (I) ~&~  Cut (II) ~&~  Cut (III) ~&~ Cut (IV)   \\      \hline     \hline
$tug $   & $17614.7 \, (\zeta_{tu})^{2}$    & $3621.8 \, (\zeta_{tu})^{2}$ & $3418.4 \, (\zeta_{tu})^{2}$ & $2650.7 \, (\zeta_{tu})^{2}$ & $1794.6 \, (\zeta_{tu})^{2}$  \\
$tcg $   & $1239.53 \, (\zeta_{tc})^{2}$    & $318.45 \, (\zeta_{tc})^{2}$ & $297.95 \, (\zeta_{tc})^{2}$ & $229.52 \, (\zeta_{tc})^{2}$ & $158.95 \, (\zeta_{tc})^{2}$  \\
$tuH$    & $14.49   \, (\eta_{tu})^{2}$     & $4.51   \, (\eta_{tu})^{2}$  & $4.08   \, (\eta_{tu})^{2}$  & $2.96   \, (\eta_{tu})^{2}$ & $1.97   \, (\eta_{tu})^{2}$  \\ 
$tcH$    & $1.649   \, (\eta_{tc})^{2}$     & $0.55   \, (\eta_{tc})^{2}$  & $0.50   \, (\eta_{tc})^{2}$ & $0.36   \, (\eta_{tc})^{2}$ & $0.24   \, (\eta_{tc})^{2}$  \\	
$tuZ$ $(\gamma_{\mu})$  & $50.56 \, (X_{tu})^{2}$ &  $14.62   \, (X_{tu})^{2}$    &  $13.09   \, (X_{tu})^{2}$  & $9.48   \, (X_{tu})^{2}$   & $6.25   \, (X_{tu})^{2}$  \\ 
$tcZ$ $(\gamma_{\mu})$  & $5.505 \, (X_{tc})^{2}$ &  $1.82   \, (X_{tc})^{2}$    &  $1.62   \, (X_{tc})^{2}$ & $1.17   \, (X_{tc})^{2}$   & $0.79   \, (X_{tc})^{2}$ \\ 				
$tuZ$ $(\sigma_{\mu\nu})$ & $81.49 \, (\kappa_{tu})^{2}$ & $22.13 \, (\kappa_{tu})^{2}$ & $20.41 \, (\kappa_{tu})^{2}$  & $15.32 \, (\kappa_{tu})^{2}$ & $10.31 \, (\kappa_{tu})^{2}$  \\
$tcZ$ $(\sigma_{\mu\nu})$ & $7.154 \, (\kappa_{tc})^{2}$ &   $2.25 \, (\kappa_{tc})^{2}$ &  $2.06 \, (\kappa_{tc})^{2}$  & $1.53 \, (\kappa_{tc})^{2}$ & $1.05 \, (\kappa_{tc})^{2}$  \\
$tu\gamma$ & $11.92 \, (\lambda_{tu})^{2}$ & $3.22 \, (\lambda_{tu})^{2}$ & $2.98  \, (\lambda_{tu})^{2}$ & $2.23   \, (\lambda_{tu})^{2}$ & $1.50   \, (\lambda_{tu})^{2}$    \\
$tc\gamma$ & $1.055 \, (\lambda_{tc})^{2}$ & $0.328\, (\lambda_{tc})^{2}$ & $0.301 \, (\lambda_{tc})^{2}$ & $0.223 \, (\lambda_{tc})^{2}$ & $0.153   \, (\lambda_{tc})^{2}$ \\			
\hline  \hline
\end{tabular}
\end{center}
\caption{ Cross-sections (in fb) of $pp \to t \bar{t} \bar{t} (\bar{t} \bar{t} t)$ with $\ell = e, \mu$ for five signal scenarios, $t q g $, $t q H$, $t q Z (\sigma_{\mu\nu})$,  $t q Z (\gamma_{\mu})$ and $t q \gamma $ passing sequential cuts explained in text.  }
\label{tab:cs}
\end{table}

As it can be seen, the three-top rate has a significant dependence on the $tqg$ FCNC coupling which 
can be understood by considering the appearance of  diagrams like $gu(c) \rightarrow t g \rightarrow t t \bar{t}$.  
The cross section for the $tqZ$ FCNC coupling are presented for both cases of  $\gamma_{\mu}$ and $\sigma_{\mu\nu}$-couplings.
Larger rate is observed for the $\sigma_{\mu\nu}$-coupling which is due to the dependence of the interaction on the $Z$ boson momentum.

To perform the whole simulation chain, \texttt{Pythia}~\cite{Sjostrand:2003wg} is used for showering and hadronization. 
Jets are reconstructed using the anti-$k_{t}$ algorithm~\cite{Cacciari:2008gp} with a cone size of 0.4.
\texttt{Delphes} framework \cite{deFavereau:2013fsa} is employed for performing a comprehensive CMS detector~\cite{Chatrchyan:2008aa}
response simulation which considers a tracker, calorimeters, and a muon system with a realistic magnetic field
configuration of the CMS detector. The $b$-tagging efficiency and  misidentification rates for light-flavor 
quarks are assumed to be dependent on the jet transverse momentum. They are
taken as~\cite{Chatrchyan:2012jua}:
\begin{eqnarray}
&&\text{$b$-tagging efficiency}~\epsilon[p_{T}] = 0.85 \tanh(0.0025 p_{T}) \Big( \frac{25}{1 + 0.063p_{T}}\Big), \nonumber \\
&&\text{misidentification rate for $c$-jets}[p_{T}]  = 0.25\tanh(0.018 p_{T}) \Big(\frac{1}{1 + 0.0013p_{T}}\Big), \\
&&\text{misidentification rate for light-jets}[p_{T}]  = 0.01 + 0.000038p_{T}\,. \nonumber
\end{eqnarray}

\begin{figure}[htb]
\begin{center}
\vspace{1cm}
\resizebox{0.45\textwidth}{!}{\includegraphics{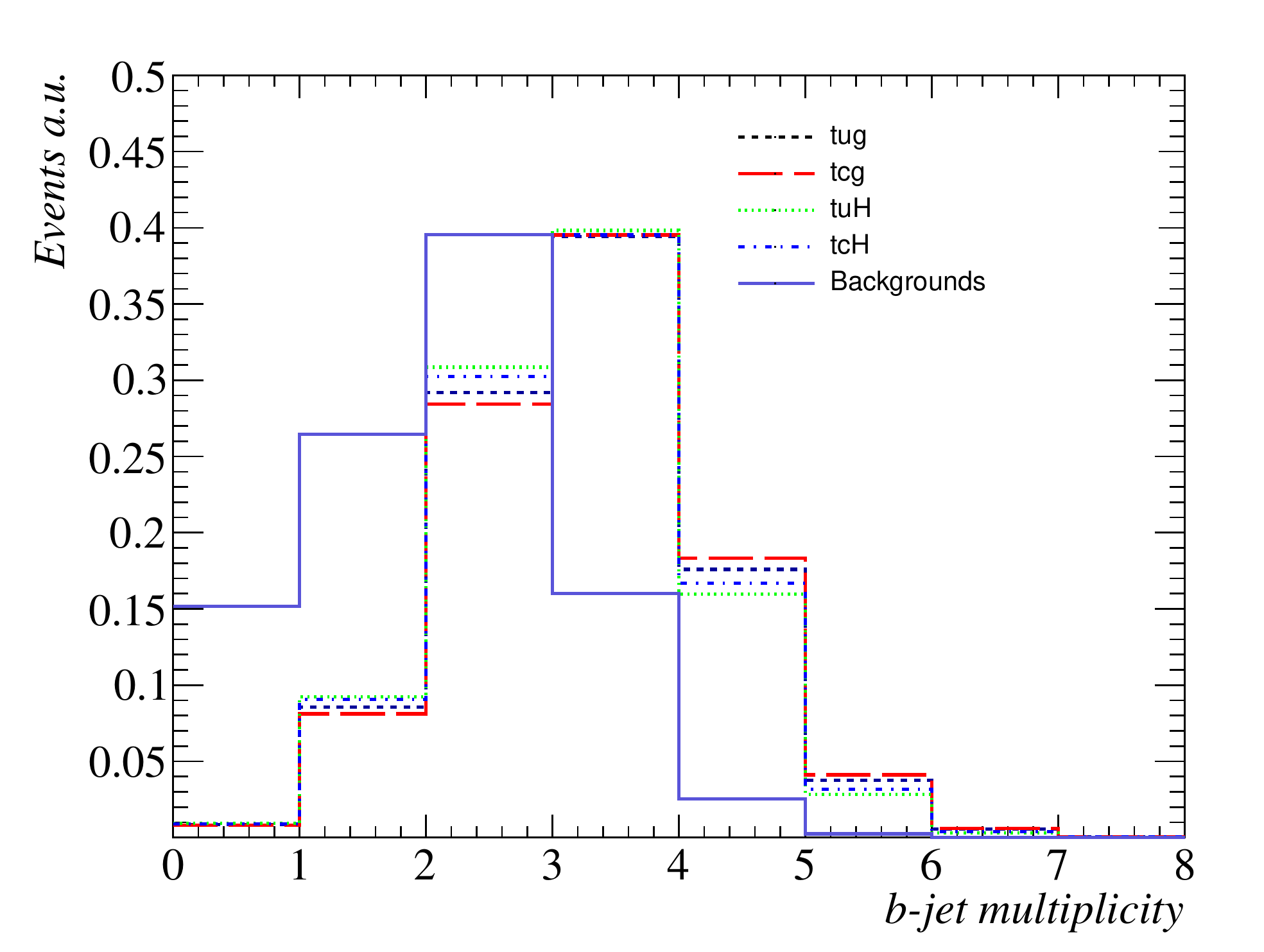}} 
\resizebox{0.45\textwidth}{!}{\includegraphics{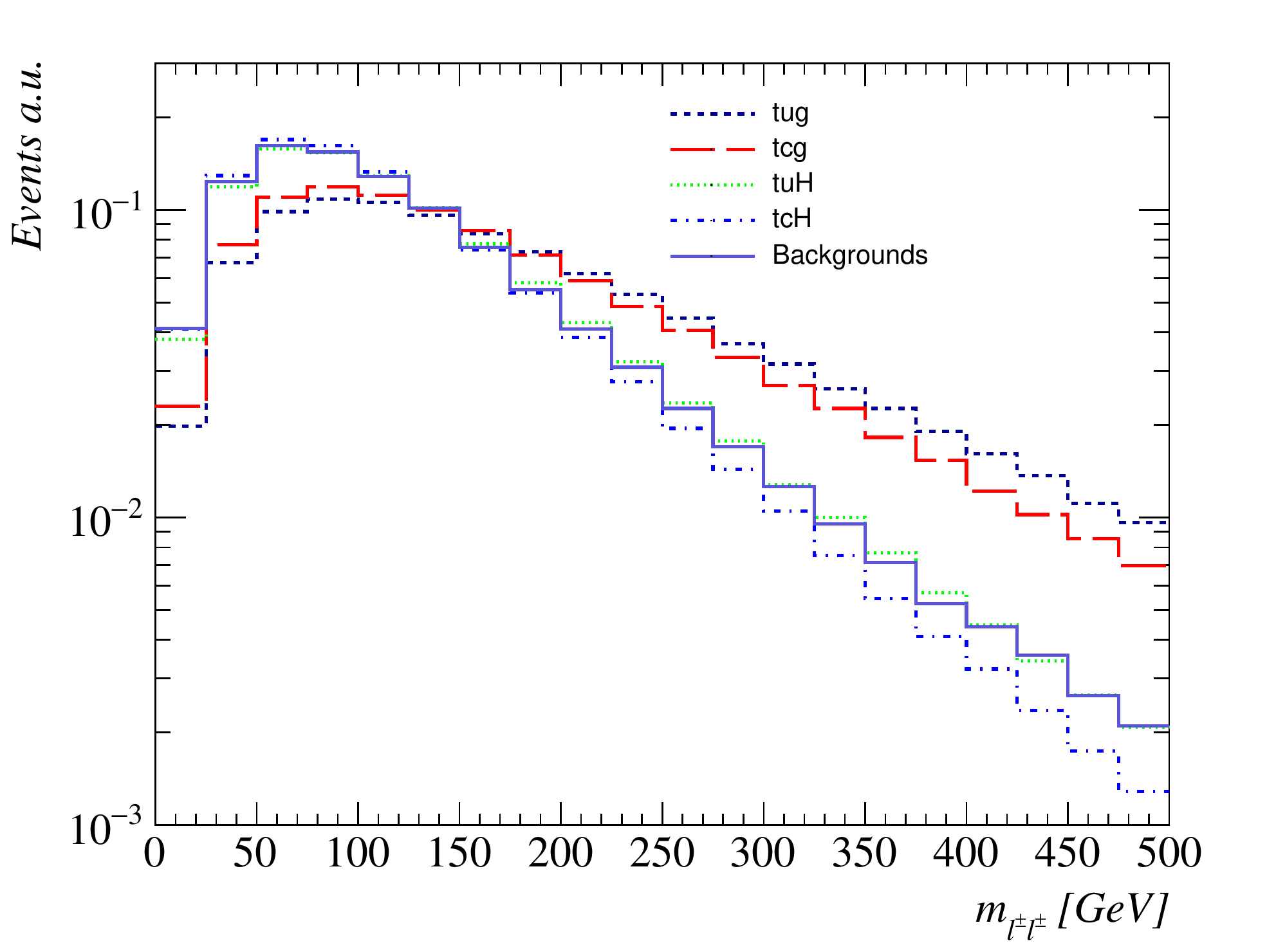}} 
\caption{ The $b$-jet multiplicity (left) and invariant mass of dilepton (right) for $tqg$ and $tqH$ signal scenarios obtained from
 \texttt{MadGraph} simulation at leading-order at $\sqrt{s} = 14$ TeV. Sum of all background processes is presented as well. }
\label{plots}
\end{center}
\end{figure}

The efficiency of $b$-tagging  for a jet with $p_{T}= 40$ GeV is $60\%$ and the misidentification rates for $c$- and light-flavor jets are 
$14\%$ and $1\%$, respectively.
The events are selected by applying the following simple criteria:
(I) Two same-sign charged lepton with $p_{T}>10~\text{GeV}, |\eta|<2.5, m_{\ell \ell}>10~\text{GeV}$,
(II) Missing Transverse Energy (MET) $>30~\text{GeV}$,
(III) At least five-jets with $p_{T}>20~\text{GeV}, |\eta_{j}|<2.5, \Delta R(\ell,j)>0.4, \Delta R(j_{1}, j_{2})>0.4$,
(IV) At least  three $b$-jets.
The $b$-jet multiplicity for the $tqg$ and $tqH$ FCNC couplings are depicted in 
the left panel of Figure~\ref{plots}. As it is expected, the FCNC signals peak at three while for the backgrounds the peak is at two.
The minimum cut of 10 GeV on the invariant mass of the same-sign dilepton is useful to reject events with pairs of same-sign energetic leptons from the heavy hadrons decays.
The invariant mass distribution of dilepton is presented in the right side of Figure~\ref{plots}.
There is a shift in the $tqg$ signal scenarios with respect to the background and the $tqH$ signals. This is due to the fact that the $tqg$ FCNC interactions are momentum 
dependent.
More kinematic cuts and complicated variables could be used to suppress the background contributions and enhance signal significance,
however including such variables is beyond the scope of this exploratory analysis and is left to a future work.
At this point, it is important to mention that the trigger of such events could be either based on only the presence of same-sign dilepton
with loose isolation requirements or based on the existence of same-sign dilepton
with lowered transverse momentum thresholds without any isolation requirement, but requiring
hadronic activities in the event~\cite{Khachatryan:2016kod}.
The cross sections of various scenarios of FCNC signals and the main background processes after
imposing different cuts are presented in Table~\ref{tab:cs} and Table~\ref{cuts}, respectively. The cross section values in both tables are given in fb unit.
The SM four-top, $t \bar{t}WW$, $t \bar{t}ZZ$, and $t\bar{t}H$ are found to have small contributions in the background composition.
Sum of the cross section of all these processes after all cuts are found to be 0.02 fb.
The three-top production in association with a jet or a $W$ boson in the SM also could contribute
to the background. After all cuts, the rate of SM three-top and $WWZ$ are found to be of the order of $10^{-3}$ fb  and  $10^{-4}$ fb which 
are neglected in this analysis.

\begin{table}[]
\centering
\caption{ \small { Cross section of the main background processes in fb after different set of cuts. } }
\label{cuts}
\begin{tabular}{c|cc} \hline\hline
Cut             &      ~~~$t \bar{t}Z$ [fb]~~~  &  ~~~$t \bar{t}W$  [fb]     ~~~ \\ \hline  \hline
 (I) $2\ell^{\pm\pm}, |\eta_{\ell}| < 2.5, p_{T} > 10$ GeV, $m_{\ell \ell} > 10$ GeV      &   1.06     &    2.66       \\ 
 (II) MET$>30$ GeV                                                                &          0.88          &    2.38       \\ 
 (III) $n_{j} \geq 5, |\eta_{j}| < 2.5,p_{T} > 20, \Delta R(\ell,j) > 0.4, \Delta R(j_{1},j_{2}) > 0.4$     &      0.48      &     0.68     \\   
 (IV) Number of $b$-jet$\geq 3$      &         0.16           &   0.21         \\
\hline
\end{tabular}
\end{table}

At this stage, we go on to set  upper limits on the signal rates at $95\%$ CL. 
In order to determine the expected limits on the FCNC branching fractions, a Bayesian approach with a flat and positive prior on the cross sections of signals are
used. The $95\%$ CL upper limits on the FCNC signal cross sections  are presented in Table~\ref{cslimits} for the integrated luminosity of
300 fb$^{-1}$.
The upper limits on signal rates are translated into the upper bounds on the FCNC branching fractions $\mathcal{B}(t\rightarrow qX)$ using the
following relations:
\begin{eqnarray}
\sigma (tug) ({\rm fb}) &=& 1869.92 ~ \mathcal{B}(t \rightarrow ug) ~,~\sigma (tcg) ({\rm fb}) = 131.58 ~ \mathcal{B}(t \rightarrow cg)\nonumber \\
\sigma (tuH) ({\rm fb}) &=& 528.83 ~ \mathcal{B}(t \rightarrow uH) ~,~ \sigma (tcH) ({\rm fb}) = 60.18 ~ \mathcal{B}(t \rightarrow cH)\nonumber \\
\sigma (tuZ)  (\gamma_{\mu}) ({\rm fb}) &=& 107.57 ~ \mathcal{B}(t \rightarrow uZ)~,~\sigma (tcZ)  (\gamma_{\mu}) ({\rm fb}) = 11.71 ~ \mathcal{B}(t \rightarrow cZ)-\gamma_{\mu} \\
\sigma (tuZ)  (\sigma_{\mu \nu}) ({\rm fb})& =& 220.24 ~ \mathcal{B}(t \rightarrow uZ) ~,~ \sigma (tcZ)  (\sigma_{\mu \nu}) ({\rm fb}) = 19.33 ~ \mathcal{B}(t \rightarrow cZ)-\sigma_{\mu \nu}\nonumber \\
\sigma (tu\gamma) ({\rm fb}) &=& 27.72 ~ \mathcal{B}(t \rightarrow tu\gamma) ~,~\sigma (tc\gamma) ({\rm fb}) = 2.45 ~ \mathcal{B}(t \rightarrow tc\gamma)	\nonumber 
\end{eqnarray}
These formulas are obtained using the functionality of the branching fractions in terms of $\zeta_{qt}$, $\eta_{qt}$, $X_{qt}$, $\kappa_{qt}$ and $\lambda_{qt}$
which could be found in Ref.\cite{saav}

\begin{table}[tbh]
\begin{center}
\begin{tabular}{c|c}
Signal    ~                          &~   Upper limit on $\sigma$ (in fb) for 300 fb$^{-1}$ ~     ~  \\      \hline     \hline
$\sigma_{95\%}(tug) $                                &    0.748  \\
$\sigma_{95\%}(tcg) $                                &   0.594    \\
$\sigma_{95\%}(tuH)$                                &   0.550       \\ 
$\sigma_{95\%}(tcH)$                                &  0.513      \\	
$\sigma_{95\%}(tuZ)$ $(\gamma_{\mu})$  &   0.616     \\ 
$\sigma_{95\%}(tcZ)$ $(\gamma_{\mu})$  &    0.530     \\ 				
$\sigma_{95\%}(tuZ)$ $(\sigma_{\mu\nu})$ &    0.602    \\
$\sigma_{95\%}(tcZ)$ $(\sigma_{\mu\nu})$ &   0.517      \\
$\sigma_{95\%}(tu\gamma)$                        &   0.605      \\
$\sigma_{95\%}(tc\gamma)$                       &    0.525      \\			
\hline  \hline
\end{tabular}
\end{center}
\caption{ The  expected  $95\%$  CL  upper  limits  on  the three top FCNC 
cross sections for five signal scenarios, $t q g $, $t q H$, $t q Z (\sigma_{\mu\nu})$,  $t q Z (\gamma_{\mu})$ and $t q \gamma $ for the integrated luminosity of
300 fb$^{-1}$.  }
\label{cslimits}
\end{table}

\begin{table}[]
\centering
\caption{ \small { The upper limits on the $tqX$ FCNC at $95\%$ CL obtained at the $\sqrt {s}$ = 14 TeV based on the integrated luminosities of  300 and 3000 fb$^{-1}$.
The HL-LHC results from a recent ATLAS experiment study which uses $t\bar{t}$ process are presented for comparison \cite{atlas-fcnc}.} }
\label{res}
\begin{tabular}{c|c|c|c} \hline\hline
Branching fraction   & three-top, 300 fb$^{-1}$   & three-top, 3 ab$^{-1}$   &     other-channels, HL-LHC, 3 ab$^{-1}$ \\ \hline\hline
$\mathcal{B}(t\rightarrow u H)$   &    $1.03 \times 10^{-3}$   &  $3.09 \times 10^{-4}$  &       $2.4\times 10^{-4}$  \cite{atlas-fcnc}   \\
$\mathcal{B}(t\rightarrow c H)$   &   $8.52 \times 10^{-3}$   &   $2.54 \times 10^{-3}$    &  $2.0\times 10^{-4}$  \cite{atlas-fcnc}  \\
$\mathcal{B}(t\rightarrow u g)$        &  $4.00 \times 10^{-4}$    &   $1.19 \times 10^{-4}$    &    -   \\
$\mathcal{B}(t\rightarrow c g)$   &    $4.51 \times 10^{-3}$   &     $1.35 \times 10^{-3}$     &     -    \\
$\mathcal{B}(t\rightarrow u Z)-\sigma_{\mu\nu}$   &  $2.73 \times 10^{-3}$    &  $8.18 \times 10^{-4}$   &   $4.3 \times 10^{-5}$     \cite{atlas-fcnc}    \\
$\mathcal{B}(t\rightarrow c Z)-\sigma_{\mu\nu}$   &   $2.67 \times 10^{-2}$   &  $7.98 \times 10^{-3}$  &     $5.8 \times 10^{-5}$     \cite{atlas-fcnc}   \\
$\mathcal{B}(t\rightarrow u Z)-\gamma_{\mu}$   &    $5.73 \times 10^{-3}$  &  $1.71 \times 10^{-3}$    &  $4.3 \times 10^{-5}$    \cite{atlas-fcnc} \\
$\mathcal{B}(t\rightarrow c Z)-\gamma_{\mu}$   &   $4.52 \times 10^{-2}$   &   $1.35 \times 10^{-2}$   &   $5.6 \times 10^{-5}$   \cite{atlas-fcnc} \\
$\mathcal{B}(t\rightarrow u \gamma)$   &   $2.18 \times 10^{-2}$    & $6.53 \times 10^{-3}$  &      $2.7 \times 10^{-5}$    \cite{CMS:2017gvo}     \\    
$\mathcal{B}(t\rightarrow c \gamma)$   &   $2.14 \times 10^{-1}$    &     $6.40 \times 10^{-2}$     &  $2.0 \times 10^{-4}$   \cite{CMS:2017gvo} \\
\hline\hline
\end{tabular}
\end{table}

The $95\%$ CL constraints on various FCNC branching fractions are summarized in Table~\ref{res} for two scenarios of integrated luminosities 300
and 3000 fb$^{-1}$ of data. The results from a recent ATLAS experiment analysis
which is based on $t\bar{t}$ process with one of the top quark decays via FCNC and another one decays in standard way
are also presented for comparison~\cite{atlas-fcnc}. As the comparison with the ATLAS limits shows, the three-top process would be able to reach 
similar sensitivity to ATLAS in the $tqH$ FCNC coupling. However, it should be noted that
the limits from the three-top process could be considerably improved taking into account the other three-top quark signatures. 
In addition, employing more powerful variables and tools (like a multivariate technique~\cite{Hocker:2007ht})
to discriminate signal events from backgrounds would lead to better sensitivities.
Taking into account the next to leading order QCD corrections to the signal processes which 
includes the three-top plus a light jet, would significantly tighten the upper bounds on the
FCNC branching fractions.

One  might  worry  for  the  validity of  the  effective field theory approach used in this analysis.  
This issue has been studies in several papers such as \cite{e1,e2,e3}.
By assuming  the  Wilson  coefficient to be equal to at most $4\pi$, one could 
translate  the bounds  on the FCNC branching fractions or equivalently the FCNC parameters
into a lower limit on the new physics scale, {\it i.e.} $\Lambda$.
For example, the upper limit on $\mathcal{B}(t \rightarrow u g)$ or correspondingly on $\zeta_{ut}^{L,R}$ leads to a lower bound
of 13.4 TeV on $\Lambda$.
Such a limit assures that the effective Lagrangian  in Eq.\ref{Eff-Lagrangian} is  valid  with  respect  to
the scale of probed momentum transfers.

\section{ Sensitivity of four-top quark production to the top quark weak and strong dipole moments }\label{Sensitivity-dipole}

\begin{figure}[htb]
\begin{center}
\vspace{0.50cm}
\resizebox{0.60\textwidth}{!}{\includegraphics{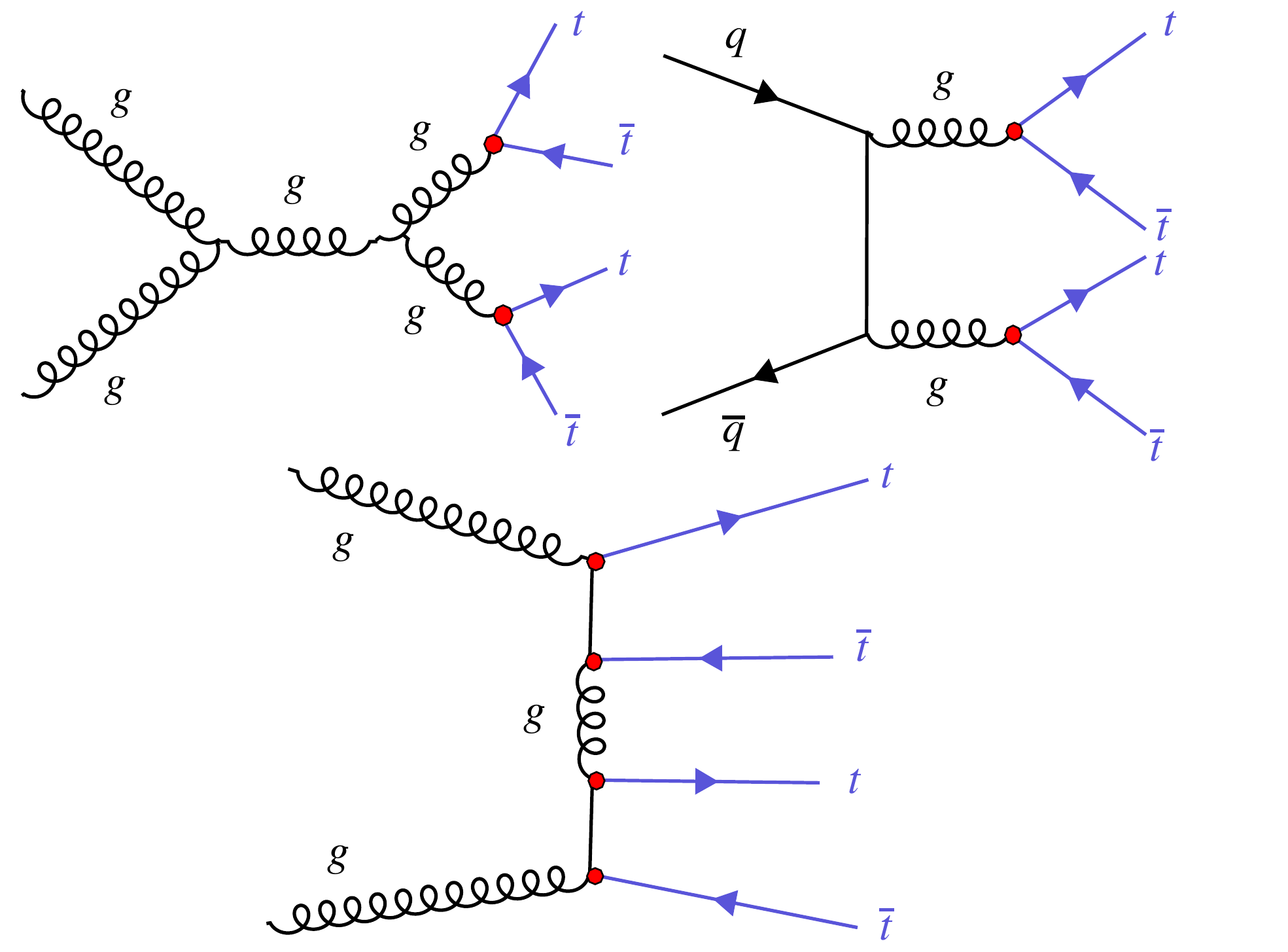}}   		
\caption{ \small { Illustrative leading order Feynman diagrams for $t \bar t t \bar t$ production representing 
the effect of strong dipole moments as filled circles.  \label{fig:Feynman-diagrams1}}  }
\end{center}
\end{figure}

In this section, we explore the sensitivity of four-top quark production in $pp$ collisions at the center-of-mass energy of 13 TeV 
to the strong ($d^{g}_{A,V}$) and weak ($d^{Z}_{A,V}$) top quark electric and magnetic dipole moments.
The representative leading order Feynman diagrams including the contributions of strong dipole moments
as filled circles are displayed in Fig.~\ref{fig:Feynman-diagrams1}.

The contributions of the top quark (weak)chromoelectric ($d_{A}^{g,Z}$), (weak)chromomagnetic ($d_{V}^{g,Z}$) dipole moments, 
coming from $O^{33}_{uW}$ and $O^{33}_{uG\phi}$ and $O^{33}_{uB\phi}$ operators, to the $tt\bar{t}\bar{t}$ production rate is determined with  the \texttt{MadGraph5\_aMC@NLO}~\cite{Alwall:2014hca}. 
By considering at most an effective vertex in each diagram, which means up to $O(\Lambda^{-2})$,  the total four-top cross section becomes at most
a quadratic function of dipole moments:

\begin{eqnarray}\label{csqq}
	\sigma(pp\rightarrow t\bar{t}t\bar{t}) (fb) &=& \sigma_{\rm SM} + 154.827 \times  d_{V}^{g} + 3404.44  \times  (d_{V}^{g} )^{2},  \nonumber \\
	\sigma(pp\rightarrow t\bar{t}t\bar{t}) (fb) &=& \sigma_{\rm SM} +2731.27  \times  (d_{A}^{g} )^{2}, \nonumber \\
	\sigma(pp\rightarrow t\bar{t}t\bar{t}) (fb) &=& \sigma_{\rm SM} - 0.689188 \times  d_{V}^{Z}  + 37.0581 \times  (d_{V}^{Z} )^{2}, \nonumber \\
	\sigma(pp\rightarrow t\bar{t}t\bar{t}) (fb) &=& \sigma_{\rm SM} + 27.962 \times  (d_{A}^{Z} )^{2},
\end{eqnarray}

where SM four-top cross section is denoted by $\sigma_{\rm SM}$ and the linear terms are the interference between the SM and new physics and its contribution 
is at $\Lambda^{-2}$ order. The quadratic terms in the cross section are corresponding to the power of $\Lambda^{-4}$
which are the first contributing terms for the strong and weak dipole moments.
The four-top cross section is symmetric with respect to $d_{A}^{g,Z} = 0$ because the cross section is a CP-even observable.
To find the coefficients of cross sections in Eq.~\eqref{csqq}, the calculations with different values of $d_{A}^{g,Z}$ and $d_{V}^{g,Z}$
are done and fit the obtained cross sections to quadratic polynomials.

In Ref.~\cite{Sirunyan:2017roi}, the CMS experiment has presented the results of a search for four-top quark production based on a dataset corresponding to an integrated luminosity
of 35.9 fb$^{-1}$ in proton-proton collisions at $\sqrt{s} = 13$ TeV.  The analysis relies on 
selecting events containing either  a same-sign lepton pair or at least three leptons (e, $\mu$) topologies. 
The observed signal significance is found to be 1.0 standard deviation and the cross section is measured to be $16.9^{+13.8}_{-11.4}$ fb which is in agreement with the standard model prediction. In Ref.~\cite{Sirunyan:2017roi}, the results are interpreted to limit the top quark Yukawa coupling ($y_{t}$) which leads to $y_{t}/y_{t}^{\rm SM} < 2.1$ at the $95\%$ confidence level.

The ATLAS experiment search for the four-top quark production is based on the single electron or muon with large transverse momentum and a high jet multiplicity topology~\cite{ATLAS:2016gqb}. The analysis has been performed using 3.2 fb$^{-1}$ of pp collisions at the center-of-mass energy of 13 TeV 
and to improve the search sensitivity, events are classified based on the jet and b-jet multiplicities. 
An upper limit of 190 fb on the four-top cross section (21 times the SM value) is set at the $95\%$ CL.  In the ATLAS study, upper bounds on the four-fermion contact interaction
and a universal extra dimensions (UED) model parameters have been set. In our study, we only use the four-top cross section measurement 
done by the CMS experiment that is used as it is the most restrictive results to date. 

The upper limits at $95\%$ CL on the strong ($d^{g}_{A,V}$) and weak ($d^{Z}_{A,V}$) dipole moments 
using the CMS experiment measurement~\cite{Sirunyan:2017roi}, which is the most recent one,  
are presented in Table~\ref{resdvda}.
The resulted bounds are compatible with the ones obtained from top quark pair cross sections at the Tevatron
and the LHC. Although the results are looser,  this study complements the capabilities of other channels at the LHC.
The four-top channel would not be able to compete with the $t\bar{t}Z$ and $t\bar{t}ZZ$ channels to probe the weak dipole moments of the top quark.

\begin{table}[]
\centering
\caption{ \small { Limits on $d^{g,Z}_{V}$ and $d^{g,Z}_{A}$ at $95\%$ CL  corresponding to current and future four-top cross section measurements. } }
\label{resdvda}
\begin{tabular}{llll} \hline\hline
Coupling    & Current four-top with 35.6 fb$^{-1}$   &  Future four-top  with 300 fb$^{-1}$  \\ \hline  \hline
$d^{g}_{V}$ &     [-0.20, 0.11]                      &      [-0.07,0.03 ]        \\
$d^{g}_{A}$ &     [-0.16, 0.16]                      &      [-0.05, 0.05]        \\ 
$d^{Z}_{V}$ &     [-1.42,1.45 ]                      &      [-0.45, 0.47]        \\
$d^{Z}_{A}$ &     [-1.65, 1.65]                      &      [-0.53, 0.53]        \\   \hline
\end{tabular}
\end{table}

%
\subsection{ Future prospect for the top quark dipole  moments }\label{sec:300fb-1}
%

To complete our study towards accessing more sensitivity to new physics effect at the LHC, 
it is important to have an estimate of the sensitivity of the four-top quark production 
using the Run-II LHC reach. In this section, we derive the limits at $95\%$ CL on the branching fractions of the
FCNC transitions and the top quark strong and weak couplings using the possible future 
reach of the LHC to measure the four-top cross section. In Ref.~\cite{Alvarez:2016nrz}, a novel strategy to search for four-top production 
based on the same-sign dilepton and the trilepton channels is presented which allows to avoid the huge backgrounds 
in the full-hadronic and mono-leptonic decay channels. Signal features such as large jet and b-jet multiplicity  
and a $Z$-mass veto in the opposite- and same-flavor dilepton spectrum, are used to suppress the main background and  increase the ratio of signal-to-background ratio as well as the signal significance. Using the suggested strategy leads to reach the following upper limit on 
the four-top signal strength using 300 fb$^{-1}$ integrated luminosity of data at $\sqrt{s} = 13$ TeV \cite{Alvarez:2016nrz}:

\begin{equation}\label{CS:300fb}
\mu_{t\bar{t}t\bar{t}} = \frac{\sigma^{\rm exp}_{t\bar{t}t\bar{t}}}{ \sigma^{\rm SM}_{t\bar{t}t\bar{t}}} < 1.87  ,
\end{equation}

where $\sigma^{\rm exp}_{tt\bar{t}\bar{t}}$ is the expected cross section of the four-top production
using 300 fb$^{-1}$ of data. Considering this estimation, we redo our analysis to determine the sensitivity of  four-top  production to the top quark dipole moments.
In Table~\ref{resdvda}, the projections of the constraints on the strong and weak dipole moments using the four-top quark
with 300 fb$^{-1}$ are depicted.  
The bounds can be translated to a lower limit on the
expected new physics scale, which is found to satisfy $\Lambda \gtrsim 4.6$ TeV. This again ensures  the validity of the
effective  Lagrangian  of  Eq.\ref{Lag:gttbar}.

We note that while the indirect constraints on $d^{g}_{V}$ and $d^{g}_{A}$, obtained 
from the rare B-meson decays and from the neutron electric dipole moment (EDM), and the direct searches
from $t\bar{t}$ events are stronger, the results of this search are complementary.
Comparison of the achieved sensitivity to  $d^{Z}_{V}$ and $d^{Z}_{A}$
with those could be obtained from the LHC and ILC presented in Section~\ref{sec:framework}, 
shows that the four-top rate does not provide competitive bounds with the $t\bar{t}Z$ channel.

%
\section{Summary and conclusions}\label{sec:Discussion}
%

Due to the very small production rate of the three- and four-top quark in the SM 
which are few fb at the LHC,
the observation of these processes is quite challenging. 
However, both the three- and four-top processes give rise to interesting set of final states 
depending on various top quarks decay modes. 
The CMS and ATLAS collaborations have measured the upper limits 
on four-top production cross section in same-sign dilepton and trilepton topologies which does not suffer from large background contributions. 
These measurements could be directly exploited to confine the NP effects. 

Uncommon decays of the top quark with extremely small rates via the flavor violating in the vertices of $tqg$, $tqZ$, $tq\gamma$, and $tqH$ are 
attracting much attention as they are considerably sensitive to several SM extensions. 
The predictions for the branching fractions of the rare decay modes  $t\rightarrow qX$, $X=g, Z, \gamma, H$ and $q=u, c$, 
in the SM are expected to be quite small so that they are unobservable at the LHC. 
However, new physics scenarios predict significant enhancement in the 
rare top quark branching fractions by many order of magnitudes.  Consequently, observation of such decay modes would indicate the existence of beyond the SM. 

In this work, we demonstrate that the three-top quark production is  sensitive to the FCNC couplings.
By performing a realistic detector simulation and the main background processes in the
same-sign dilepton channel, upper limits on the FCNC branching fractions are obtained. 
Upper limits of the order of $10^{-4}$ at $95\%$ CL on the branching fractions of $t \rightarrow qH$, $t \rightarrow qg$ and $t \rightarrow qZ$ are set.
We also examine the sensitivity of the four-top cross section to the strong and weak top quark dipole moments. Upper limits are set on 
the top quark dipole moments which are compatible with the other bounds extracted from the other processes at the LHC.

%
\section*{Acknowledgments}
%

Malihe Malekhoseini and Mehrdad Ghominejad thank the University of Semnan for financial support of this project.
Hamzeh Khanpour is thankful to the University of Science and Technology of Mazandaran, and School of Particles and Accelerators, Institute for Research in Fundamental Sciences (IPM) for financial support provided for this research. M. Mohammadi Najafabadi would like to appreciate the Iran National Science Foundation (INSF) for the financial support.


%
%


%

\end{document}